\documentclass[a4paper,11pt]{article}
\usepackage{pos}

\usepackage{caption,subcaption}
\usepackage{wrapfig}

\usepackage{siunitx}

\usepackage{cleveref}

\usepackage[framemethod=tikz]{mdframed} % Box off stuff
\usepackage{color} % Colour support
\newcommand{\MyArrow}[4]{
        \draw[line width=0.5pt] (#1,#2) -- (#3,#4); %#1+#3/2-#1/2,#2+#4/2-#2/2
        \draw [-stealth, line width=0.5pt](#1,#2) -- (#1+#3/2-#1/2,#2+#4/2-#2/2);
        }

\title{Partial-wave decomposition of the diffractively produced $\pi^-\pi^+\pi^-\eta$ final state at COMPASS}
\ShortTitle{Partial-wave decomposition of the diff. produced $\pi^-\pi^+\pi^-\eta$ final state at COMPASS}

\author*[a]{David Spülbeck}
\onbehalf{for the COMPASS Collaboration}

\affiliation[a]{Helmholtz-Institut für Strahlen- und Kernphysik, Bonn University, Germany}

\emailAdd{spulbeck@cern.ch}

\abstract{
One of the prime goals of the COMPASS experiment at CERN is the study of the light meson spectrum, with a particular emphasis on the search for exotic states. We present the first high-statistics partial-wave decomposition of the diffractive reaction $\pi^-+p\to \pi^-\pi^+\pi^-\eta+p$, which spans a wide range of decay channels, such as $f_1(1285)\pi^-$, $a_2^-(1320)\eta$, $\eta^\prime\pi^-$, and $\rho(770)a_0^-(980)$. 
This analysis also includes decay channels, i.e. $f_1(1285)\pi^-$ and $\eta^\prime\pi^-$, predicted by theoretical models for the lightest hybrid meson and providing the opportunity to verify the hybrid meson hypothesis of the $\pi_1(1600)$.
}

\FullConference{The 21st International Conference on Hadron Spectroscopy and Structure (HADRON2025)\\
 27 - 31 March, 2025\\
Osaka University, Japan\\}

\begin{document}
\maketitle

% \nopagebreak
% \vspace{-2ex}
\section{Introduction}

As part of the COMPASS spectroscopy programme, diffractive dissociation of $\pi^-$ beams was studied using a 190~GeV/$c$ hadron beam provided by CERN M2 beam line \cite{COMPASS_hadron}. The largest data set was recorded with a liquid hydrogen target. At these high energies, the process is dominated by diffractive production, which predominantly produces excited isovector mesons such as $\pi_J$ and $a_J$. This $t$-channel scattering occurs at low momentum transfer, characterized by the reduced momentum transfer variable $t^\prime = |t| - |t|_\text{min}$. With this setup, COMPASS has collected the world's largest data set for such diffractive processes, enabling detailed studies of light isovector mesons. Our analyses of systems such as $(3\pi)^-$, $\eta^{(\prime)}\pi^-$, $\omega\pi\pi$, and $K_S K$ \cite{LightIsoVector,OddAndEven,OmegaPiPi,KsK} have played a key role in improving the precision of resonance parameter measurements and clarifying the complex light-meson spectrum, for instance, in the cases of the $a_1(1420)$ \cite{triangle} and the spin-exotic $\pi_1(1600)$ \cite{poleofpione}.

The $\pi^-\pi^+\pi^-\eta$ system is particularly compelling due to its four-body nature, providing access to decay channels that have not been studied previously with COMPASS data, especially those involving isobars in the $\pi^+\pi^-\eta$ subsystem. Moreover, this system is predicted by several theoretical models to contain multiple decay modes of the lightest hybrid meson candidate \cite{dudek}, the $\pi_1(1600)$, making it a promising channel for investigating hybrid meson dynamics.

We reconstruct the $\pi^-\pi^+\pi^-\eta$ system by selecting events with three charged tracks originating from the primary interaction vertex within the target volume and exactly two photons consistent with an $\eta$ meson decay. For the photon pair, a kinematic fit to the nominal $\eta$ mass is performed, and a cut on the fit confidence level is applied. After applying all standard selection criteria, we retain approximately 745,000 exclusive events within the kinematic range $m_{3\pi\eta} < 3~\text{GeV}/c^2$ and $0.1~(\text{GeV}/c)^2 \leq t^\prime \leq 1.0~(\text{GeV}/c)^2$, which are then used in the partial-wave analysis (PWA). Up to date, the most accurate analysis of the $\pi^-\pi^+\pi^-\eta$ system was done by the BNL E852 collaboration based on 89,000 events \cite{BNL_F1}.

% Notes: 
% \begin{itemize}
%     \item Thehadron program studying diff. pion diss.
%     \begin{itemize}
%         \item beam momentum, reggeon exchange, single pommeron  exchange dominant
%         \item t-channel reaction $t^\prime$
%         \item worlds largest data set allows not just for precision measurements, but also to tidy up the overpopulated experimentally claimed spectrum of iso-vector mesons.  
%     \end{itemize}
%     \item Good playground to study exotic mesons, especially spin-exotics: pi1(1600)
%     \item analyses of different systems already helped to clarify open questions on several states (a1(1420), pi1->eta(')pi/rho(770)pi, omega pi pi)
%     \item First analysis of 3pieta gives access to new decay channels and to dominant decay channels of the pi1(1600) in the same system, which is experimentally interesting as branching ratios can be estimated 
%     \item Event selection: three tracks and two photons in Ecals which are kinematically fitted to the nominal eta mass with CL cut. Finally, 700k exclusive events
% \end{itemize}

% \nopagebreak
% \vspace{-2ex}
\section{Partial-wave decomposition of \texorpdfstring{$\pi^-\pi^+\pi^-\eta$}{pipipieta}}

% Improved version:
The goal of the partial-wave decomposition is to decompose the data set into partial-wave amplitudes with well-defined quantum numbers and decay channels. Each partial wave is uniquely labeled by $J^{PC}M^\epsilon I_1I_2 LS$, where $J^{PC}$ denotes the total spin, parity, and charge conjugation quantum numbers of the intermediate state, $M$ is the spin projection, and $\epsilon=\pm1$ indicates the naturality of the exchanged object (often referred to as a Reggeon in Regge theory). The indices $I_1$ and $I_2$ indicate the first two daughter mesons, also referred to as isobars, of the decay channel, and $L$ and $S$ refer to the relative orbital angular momentum and spin between them, respectively.

The partial-wave amplitudes are extracted in bins of the invariant mass $m_{3\pi\eta}$ and $t^\prime$ using extended maximum likelihood fits. This allows for a model-independent determination of the amplitude structure as a function of both $m_{3\pi\eta}$ and $t^\prime$, which can subsequently be interpreted in a resonance-model fit. The high statistics available at COMPASS enable these fits to be performed in narrow $t^\prime$-bins, which is essential for distinguishing resonant contributions from non-resonant coherent background, due to their different $t^\prime$-dependencies. We choose a bin width of \SI{40}{MeV/c^2} for $m_{3\pi\eta}$. For $t^\prime$, we first divide the data into three $t^\prime$ bins of equal number of events. The bin with the highest $t^\prime$ range is subsequently split into two, yielding four $t^\prime$ bins in total.

The intensity distribution for an event with phase-space variables $\tau=\{\tau_1,\tau_2\}$, where $\tau_1$ and $\tau_2$ correspond to the two possible $\pi^-$ combinations, is modeled as a sum of incoherent contributions. Each of these incoherent terms contains a coherent sum over a subset of partial waves $i=(a,b)$:

\begin{align}
\mathcal{I}(\tau=\{\tau_1,\tau_2\})=
\underbrace{\left|\sum\limits_{a}\mathcal{T}_{a,r}^{\epsilon}{}^{\text{sy}}\Psi_a^{\epsilon}(\tau)\right|^2}_{\text{non. $\eta^\prime\pi^-$-waves}}+\underbrace{\left|\sum\limits_{b}\mathcal{T}_{b,r}^{\epsilon}\Psi_b^{\epsilon}(\tau_1)\right|^2 +\left|\sum\limits_{b}\mathcal{T}_{b,r}^{\epsilon}\Psi_b^{\epsilon}(\tau_2)\right|^2}_{\text{$\eta^\prime\pi^-$-waves}} +\left|\mathcal{T_{\text{flat}}}\right|^2.
\end{align}
Here, $b$ refers to the $\eta^\prime\pi^-$ partial waves, and $a$ refers to all others. The decay amplitudes for the non-$\eta^\prime\pi^-$ waves are symmetrized under the exchange of the two identical $\pi^-$ mesons. Since the $\eta^\prime$ has a very narrow width (\SI{188}{keV/c^2}), the interference between waves of kind $a$ and $b$, and between the two $\pi^-$ combinations ($\tau_1$ and $\tau_2$) is negligible, justifying their incoherent treatment. A flat wave is included to account for non-resonant background uniformly distributed over phase space.

Each partial wave amplitude is factorized into a decay amplitude $\Psi_i$, computed in the isobar model, and a complex-valued transition amplitude $\mathcal{T}_i$, which encodes its strength and relative phase. These transition amplitudes are free parameters in the fit.

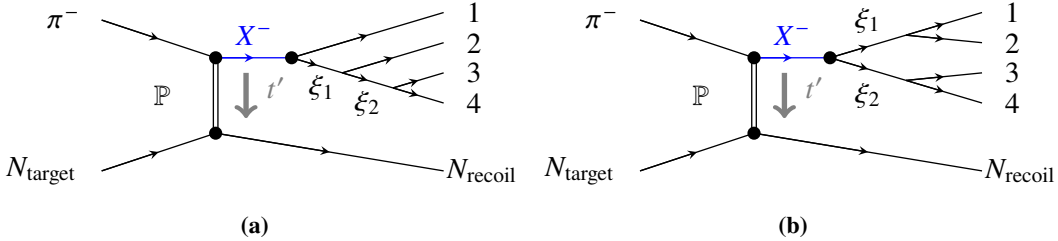
\begin{figure}[ht]
    \centering
    \subfloat[\centering ]{
        \centering
        \begin{tikzpicture}[scale=1.0, transform shape]
            % Beam
            \MyArrow{-1.5}{0.5}{0}{0};
            \node[align=left] at (-2.0,0.5) {$\pi^-$};
            % X
            \draw[blue,{-stealth}, line width=0.5pt](0,0) -- (0.5,0);
            \draw[blue,line width=0.5pt] (0,0) -- (1,0);
            \node[align=left] at (0.5,0.3) {\textcolor{blue}{$X^-$}};
            % Final State
            \MyArrow{1}{0}{3}{0.6};
            \node[align=left] at (3.4,0.6) {1};
            \MyArrow{1}{0}{1.667}{-0.2};
            \node[align=left] at (1.4,-0.4) {$\xi_1$};
            \MyArrow{1.667}{-0.2}{3}{0.2};
            \node[align=left] at (3.4,0.2) {2};
            \MyArrow{1.667}{-0.2}{2.334}{-0.4};
            \node[align=left] at (2.0,-0.6) {$\xi_2$};
            \MyArrow{2.334}{-0.4}{3}{-0.2};
            \node[align=left] at (3.4,-0.2) {3};
            \MyArrow{2.334}{-0.4}{3}{-0.6};
            \node[align=left] at (3.4,-0.6) {4};
            % Pommeron
            \draw[line width=0.5pt] (-0.03,-1) -- (-0.03,0);
            \draw[line width=0.5pt] (0.03,-1) -- (0.03,0);
            \node[align=left] at (-0.7,-0.5) {$\mathbb{P}$};
            % tprime
            \draw[gray, -To, line width=2pt](0.4,-0.2) -- (0.4,-0.8);
            \node[gray, align=left] at (0.8,-0.4) {\textbf{$t^\prime$}};
            % Target
            \MyArrow{-1.5}{-1.5}{0}{-1};
            \node[align=left] at (-2.3,-1.5) {$N_{\text{target}}$};
            % Recoil
            \MyArrow{0}{-1}{3}{-1.5};
            \node[align=left] at (3.5,-1.5) {$N_{\text{recoil}}$};
            % nods
            \fill[black] (0,0) circle (0.5ex);
            \fill[black] (0,-1) circle (0.5ex);
            \fill[black] (1,0) circle (0.5ex);
        \end{tikzpicture}}
    \subfloat[\centering ]{
        \centering
        \begin{tikzpicture}[scale=1.0, transform shape]
            % Beam
            \MyArrow{-1.5}{0.5}{0}{0};
            \node[align=left] at (-2.0,0.5) {$\pi^-$};
            % X
            \draw[blue,{-stealth}, line width=0.5pt](0,0) -- (0.5,0);
            \draw[blue,line width=0.5pt] (0,0) -- (1,0);
            \node[align=left] at (0.5,0.3) {\textcolor{blue}{$X^-$}};
            % Final State
            \MyArrow{1}{0}{2}{0.3};
            \node[align=left] at (1.5,0.5) {$\xi_1$};
            \MyArrow{2}{0.3}{3}{0.6};
            \node[align=left] at (3.4,0.6) {1};
            \MyArrow{2}{0.3}{3}{0.2};
            \node[align=left] at (3.4,0.2) {2};
            \MyArrow{1}{0}{2}{-0.3};
            \node[align=left] at (1.5,-0.5) {$\xi_2$};
            \MyArrow{2}{-0.3}{3}{-0.2};
            \node[align=left] at (3.4,-0.2) {3};
            \MyArrow{2}{-0.3}{3}{-0.6};
            \node[align=left] at (3.4,-0.6) {4};
            % Pommeron
            \draw[line width=0.5pt] (-0.03,-1) -- (-0.03,0);
            \draw[line width=0.5pt] (0.03,-1) -- (0.03,0);
            \node[align=left] at (-0.7,-0.5) {$\mathbb{P}$};
            % tprime
            \draw[gray, -To, line width=2pt](0.4,-0.2) -- (0.4,-0.8);
            \node[gray, align=left] at (0.8,-0.4) {\textbf{$t^\prime$}};
            % Target
            \MyArrow{-1.5}{-1.5}{0}{-1};
            \node[align=left] at (-2.3,-1.5) {$N_{\text{target}}$};
            % Recoil
            \MyArrow{0}{-1}{3}{-1.5};
            \node[align=left] at (3.5,-1.5) {$N_{\text{recoil}}$};
            % nods
            \fill[black] (0,0) circle (0.5ex);
            \fill[black] (0,-1) circle (0.5ex);
            \fill[black] (1,0) circle (0.5ex);
        \end{tikzpicture}}
     \caption{Two possible event topologies taken into account.}
    \label{fig:topologies}
\end{figure}

\begin{figure}[ht]
        \hbox{\subfloat[\centering ]{\includegraphics[width=0.29\textwidth]{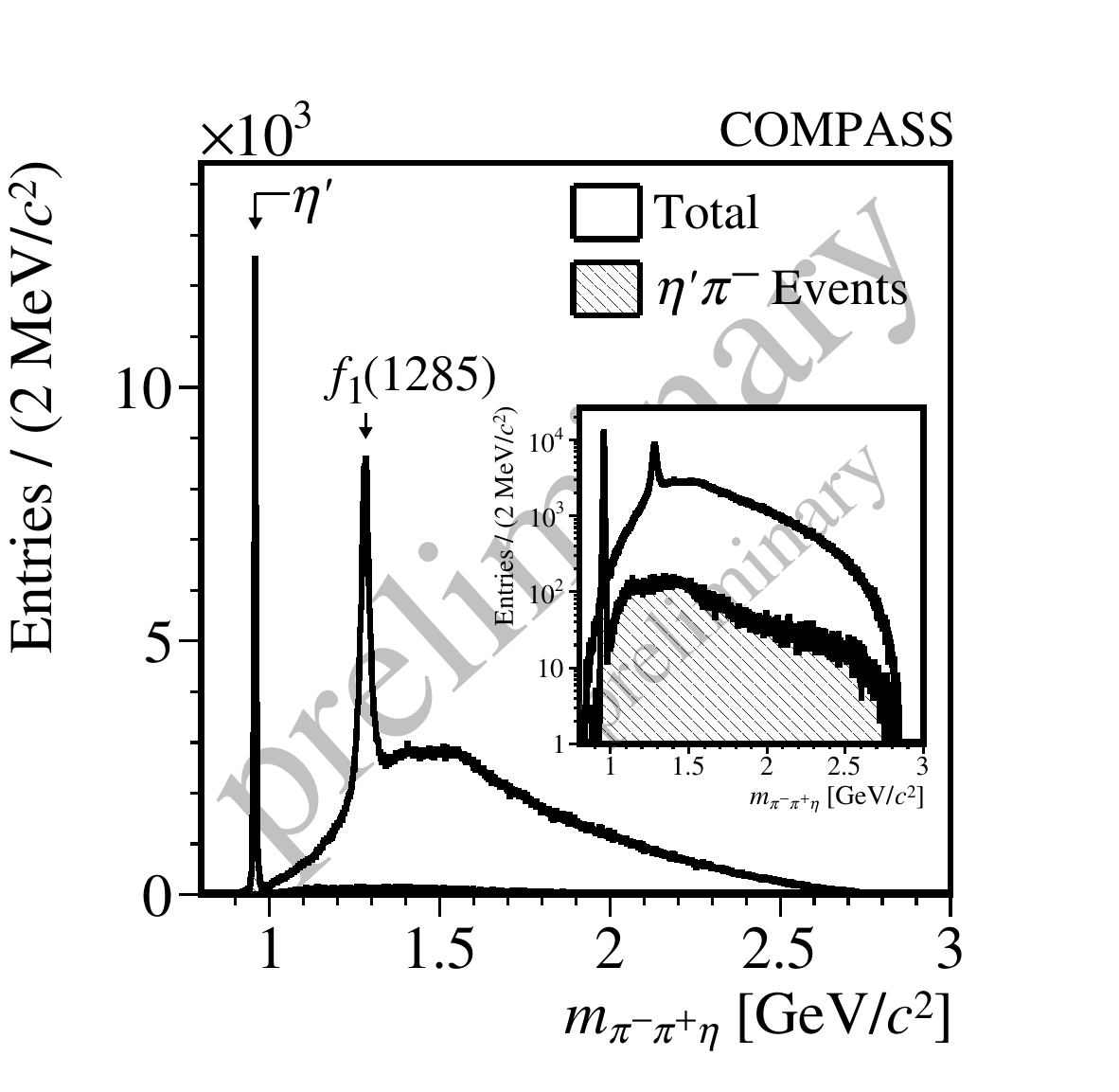}}\hspace{-0.75cm}
        \subfloat[\centering ]{\includegraphics[width=0.29\textwidth]{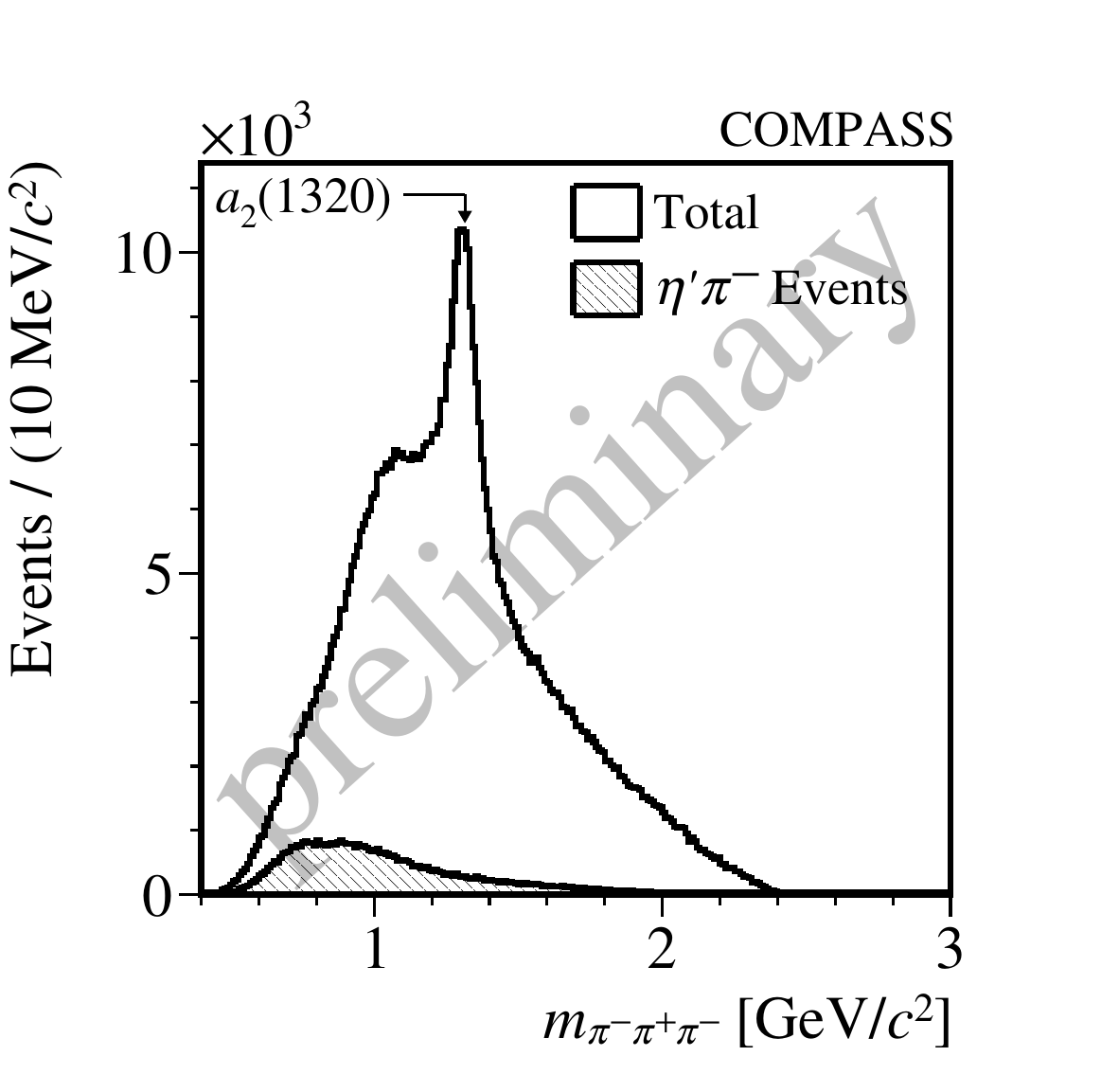}}\hspace{-0.75cm}
        \subfloat[\centering ]{\includegraphics[width=0.29\textwidth]{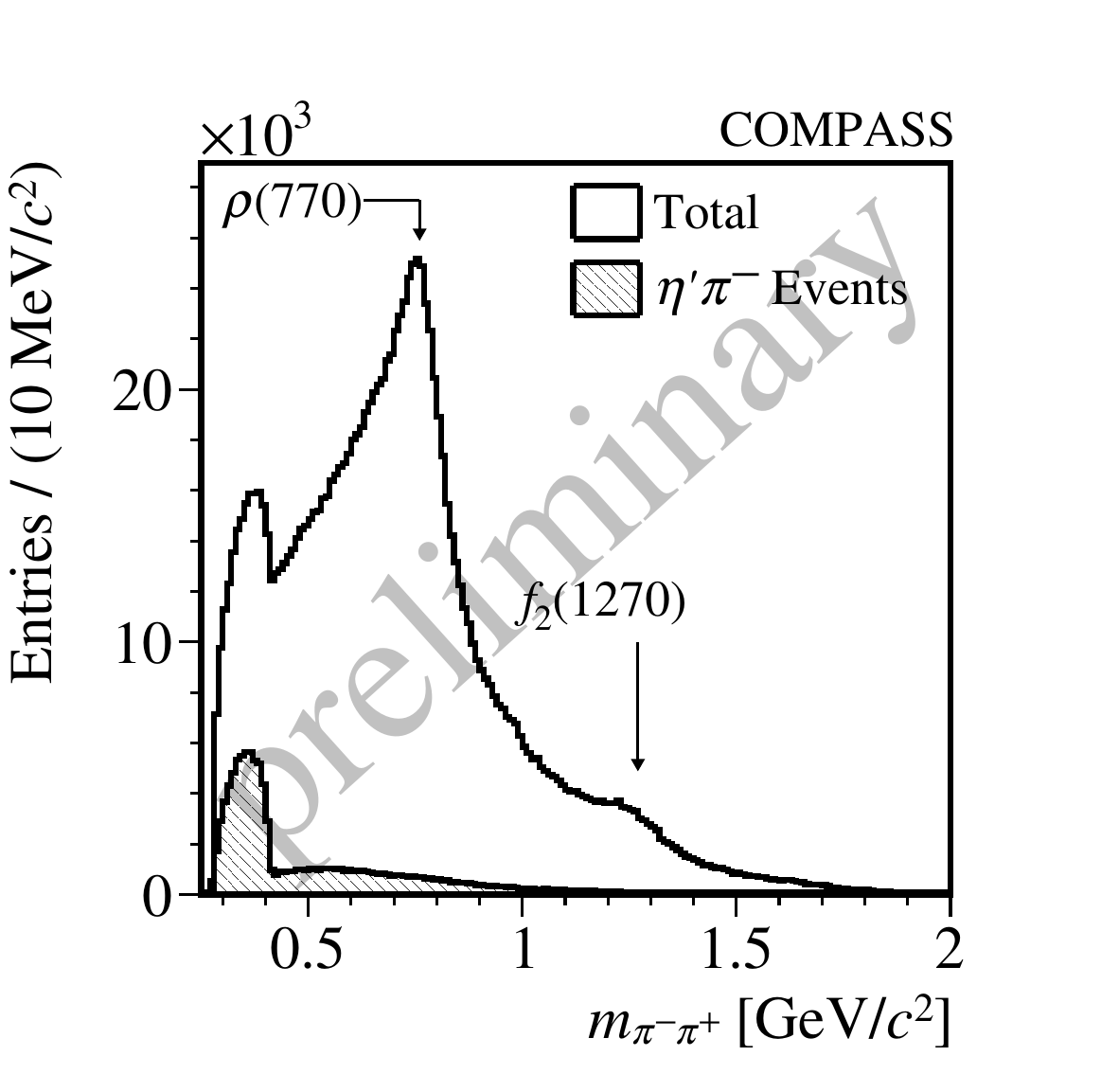}}\hspace{-0.75cm}
        \subfloat[\centering ]{\includegraphics[width=0.29\textwidth]{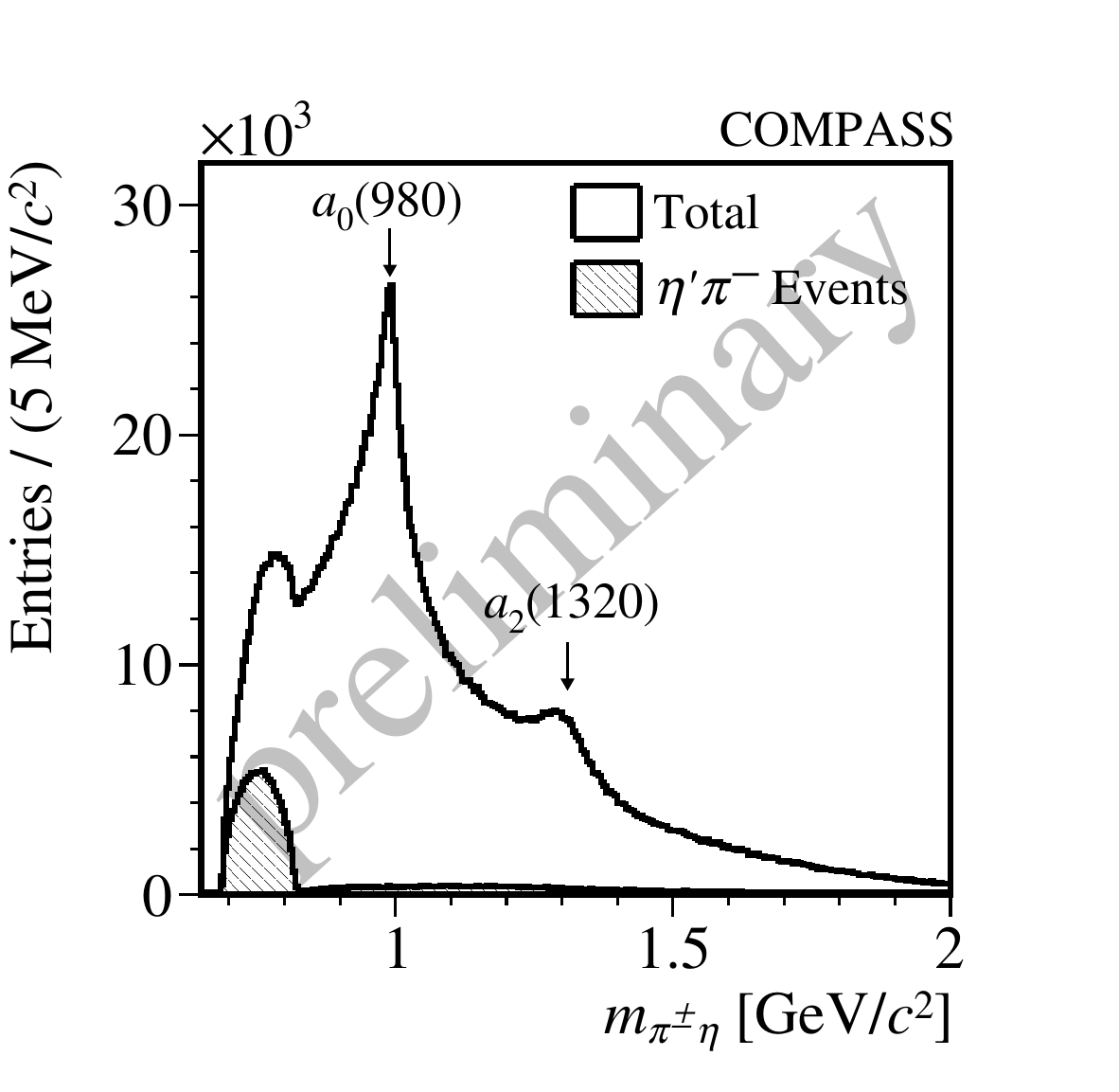}}}
        \qquad
        \caption{Invariant mass distribution of the subsystems: (a) $\pi^-\pi^+\eta$, (b) $\pi^-\pi^+\pi^-$, (c) $\pi^-\pi^+$, and (d) $\pi^\pm\eta$. The dominant and as a peak appearing isobars are highlighted. The hashed distribution shows those events where at least one $\pi^-\pi^+\eta$ combination is within the $\eta^\prime$-peak.}
        \label{fig:subsystems}
\end{figure}

In the isobar model the decay of the excited state $X^-$ into the four-body final state is described as a sequence of two-body decays. Following this ansatz, we take two possible decay topologies into account depicted in Fig.~\ref{fig:topologies} a) and b). In addition, a third topology is considered dedicated to partial waves of kind $b$ where the $\eta^\prime$ decays via a three-body decay into $\pi^-\pi^+\eta$. This decay is modeled by the amplitude discussed in \cite{Kubis}.We show the invariant mass distributions of the subsystems $\pi^+\pi^-\eta$, $\pi^-\pi^+\pi^-$, $\pi^-\pi^+$, and $\pi^\pm\eta$ with prominent isobars highlighted in Fig.~\ref{fig:subsystems}. In total, our model includes 15 isobars.

To identify a wave-set that describes the data best, we perform wave-set selection fits. The procedure is as follows: i) We truncate the partial-wave expansion by requiring $J,L$\,<\,$7$ and $M$\,<\,$2$, resulting in 488 partial waves plus the flat wave. ii) For each wave, we determine a threshold in $m_{3\pi\eta}$ above which the phase-space of the wave becomes non negligible and we start to consider it in the fit. iii) We perform wave-set selection fits by maximizing a modified likelihood function that includes a Cauchy regularization \cite{KsK} term for each partial wave. This term penalizes insignificant amplitudes, pulling them toward zero. Furthermore, we fit over seven neighboring ($m_{3\pi\eta},t^\prime$)-bins simultaneously, constraining the transition amplitudes of a given wave to vary smoothly across bins. This regularization improves the fit's stability and enforces continuity across bins. The final wave set is individual for each ($m_{3\pi\eta},t^\prime$)-bin. In total, 290 distinct partial waves plus the flat wave are considered across all bins, with the bin containing the largest number of waves including 106. Using these wave sets, we perform the main fit without the regularization terms. For each bin, 100 fits are performed with randomly chosen initial parameter values to scan the parameter space for a global minimum. In the range, $m_{3\pi\eta}<$\SI{2.5}{GeV/c^2}, the fit converges to the same minimum in nearly all of the trials. In a few bins, this convergence efficiency decreases slightly, dropping to around 50\%. In the range, $m_{3\pi\eta}>$\SI{2.5}{GeV/c^2}, the fit stability drops; in the least stable bin, only about 10\% of the fits converge to the same minimum. More details about the PWA can be found in section 5 of Ref.~\cite{Ketzer}.

\section{Results}

\begin{figure}[!ht]
\centering
    %\subfloat[\centering ]{\includegraphics[width=0.33\textwidth]{figures/Reflectivities.pdf}}
    \subfloat[\centering ]{\includegraphics[width=0.45\textwidth]{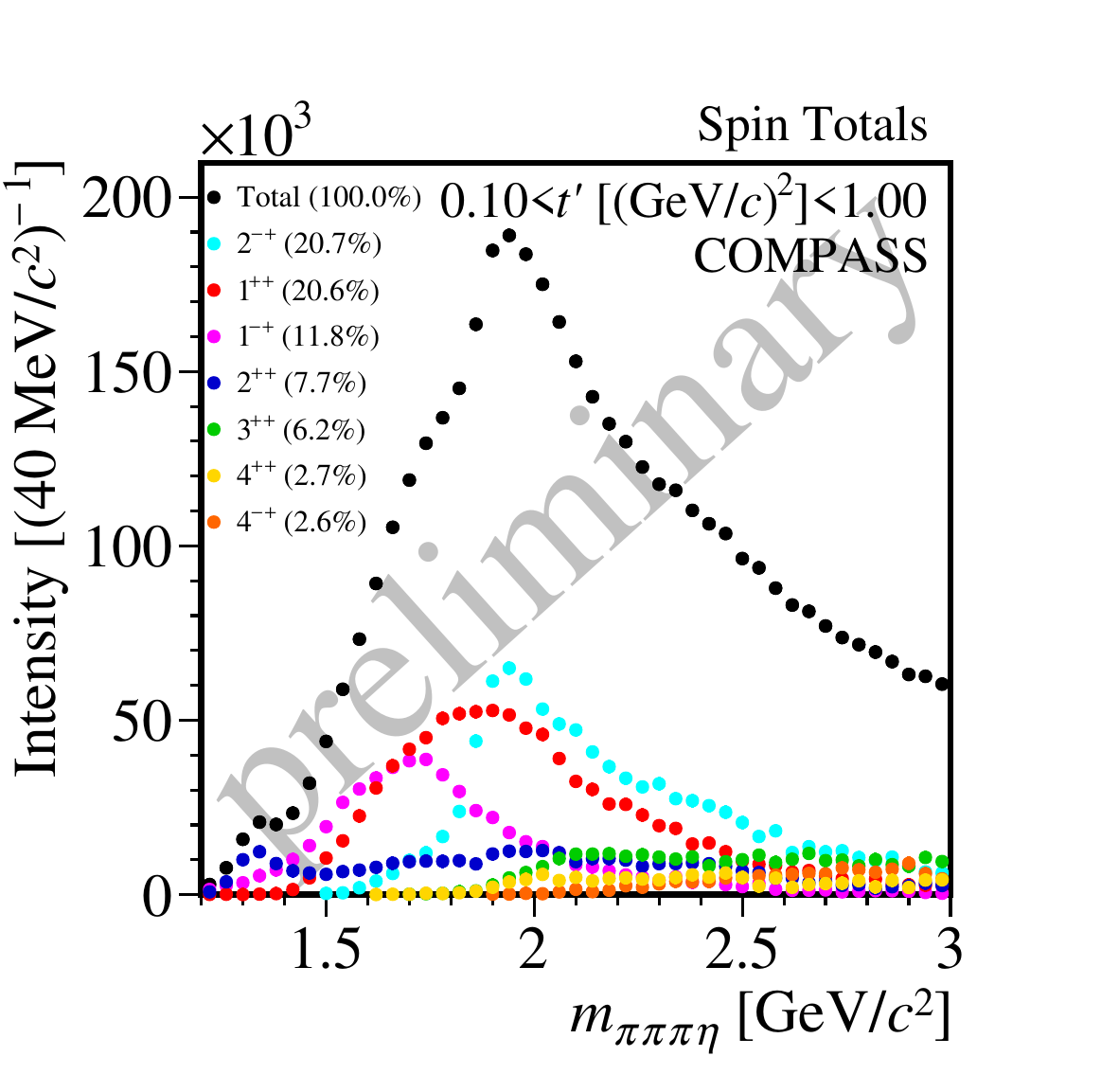}}
    \subfloat[\centering ]{\includegraphics[width=0.45\textwidth]{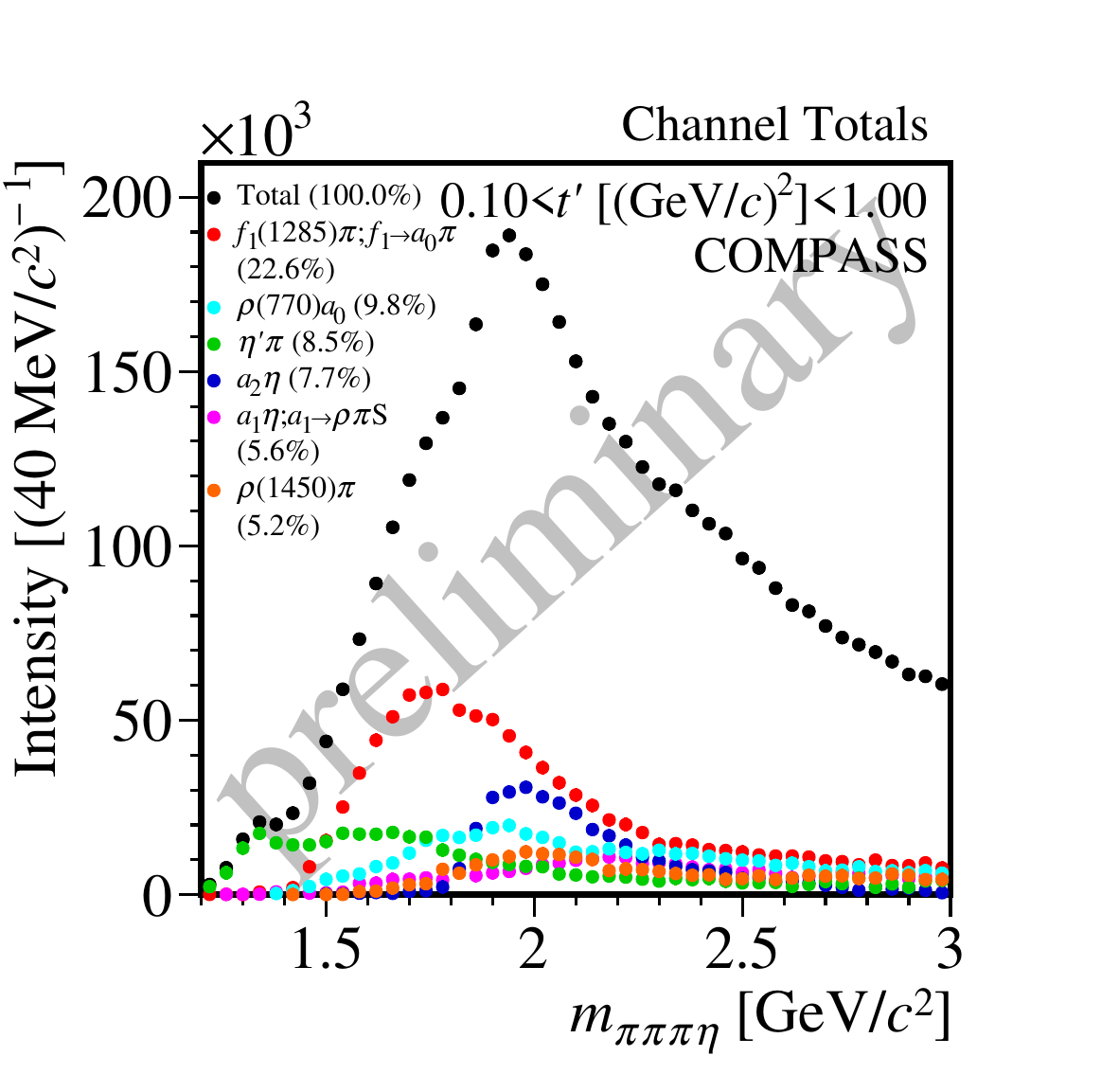}}
    \caption{%(a) Shows the full intensity, the contribution of all partial-waves, and the flat wave summed over all $t^\prime$-bins. 
   (a) Full intensity and the dominating $J^{PC}$-sectors summed over all $t^\prime$-bins. (b) The same as in (a) but for the dominant decay channels. }%
    \label{fig:Totals_Large_TpSum}%
\end{figure}

To provide a general overview of the results, we show two levels of decomposition in Fig.~\ref{fig:Totals_Large_TpSum}: the coherent contribution of each spin-parity sector (a) and the coherent contribution of individual decay channels (b). %The flat wave accounts for approximately one quarter of the total intensity and exhibits a smooth, broad distribution. Its relative contribution increases with higher invariant masses—an expected trend, as the fit model primarily describes diffractive production, which dominates at lower masses. A similar behavior of the flat wave has been reported in~\cite{BNL_F1}.
%Fig.~\ref{fig:Totals_Large_TpSum} (b) shows the total intensity (black) along with the contributions from the most prominent spin-parity sectors (colored curves). 
Consistent with observations in other systems studied at COMPASS, the dominant sectors are $J^{PC}=1^{++}$, $2^{-+}$, $2^{++}$, and $3^{++}$. Notably, the spin-exotic $J^{PC}=1^{-+}$ sector contributes significantly, with a relative intensity fraction of 11.8\%. This highlights the $\pi^-\pi^+\pi^-\eta$ system as a promising environment for investigating the lightest hybrid meson candidate, the $\pi_1(1600)$. The most prominent contributions arise from the $f_1(1285)\pi$, $\rho(770) a_0$, $\eta^\prime\pi$, and $a_2\eta$ channels.

% Notes
    % \begin{itemize}
    %     \item Total, Partial-waves, Flat wave 
    %     \item Spin totals
    %     \item Channel totals
    %     \item Comment on dominant sectors and channels
    % \end{itemize}

%Resonance-like structures are identified through the analysis of intensity distributions of individual partial waves and the relative phases. 
A characteristic signature of a resonance is the simultaneous observation of a peak in the intensity and a rapid variation in the relative phase. Applying this criterion, we identify more than 15 single waves exhibiting such features. The selected waves span quantum numbers $J^{PC} = (1,2,3,4)^{++}$ and $(1,2)^{-+}$ across various decay channels, providing the opportunity to investigate both new decay modes of established resonances and potentially new states, such as in the $a_3$ sector.

\begin{figure}[!ht]
        \centering
        \hbox{\subfloat[\centering ]{\includegraphics[width=0.38\textwidth]{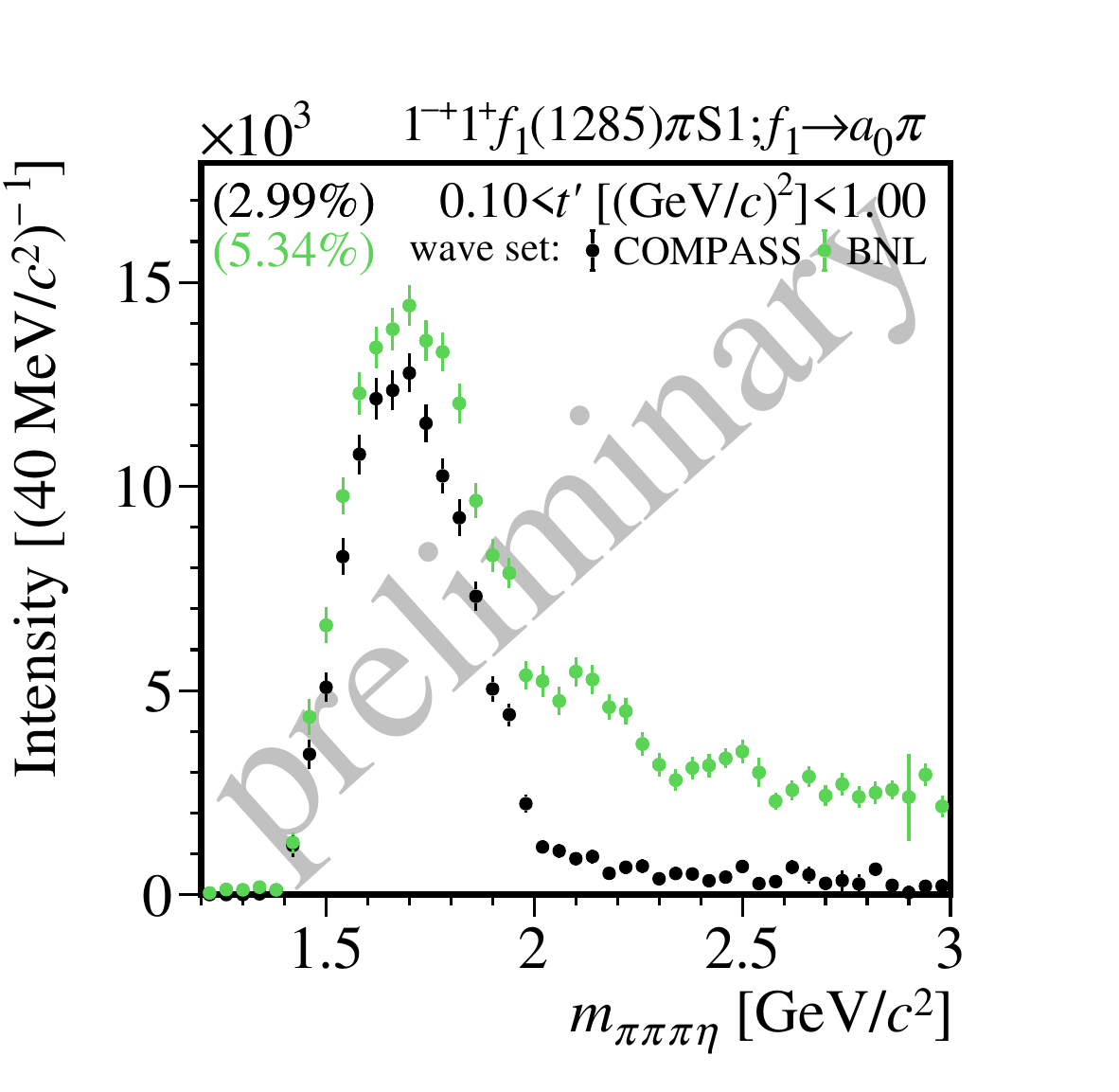}}
        \hspace{-0.8cm}
        \subfloat[\centering ]{\includegraphics[width=0.38\textwidth]{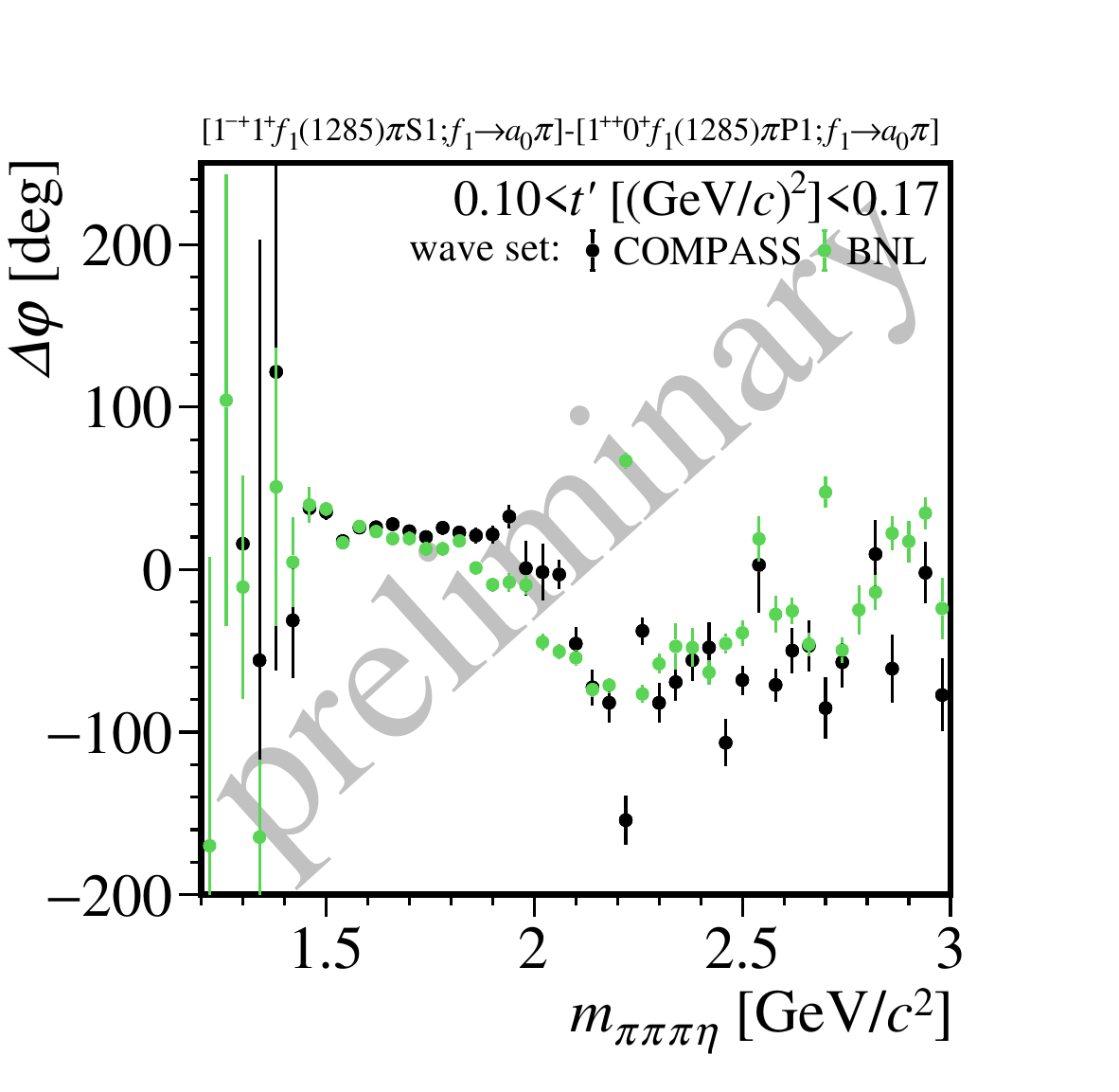}}
        \hspace{-0.8cm}
        \subfloat[\centering ]{\includegraphics[width=0.38\textwidth]{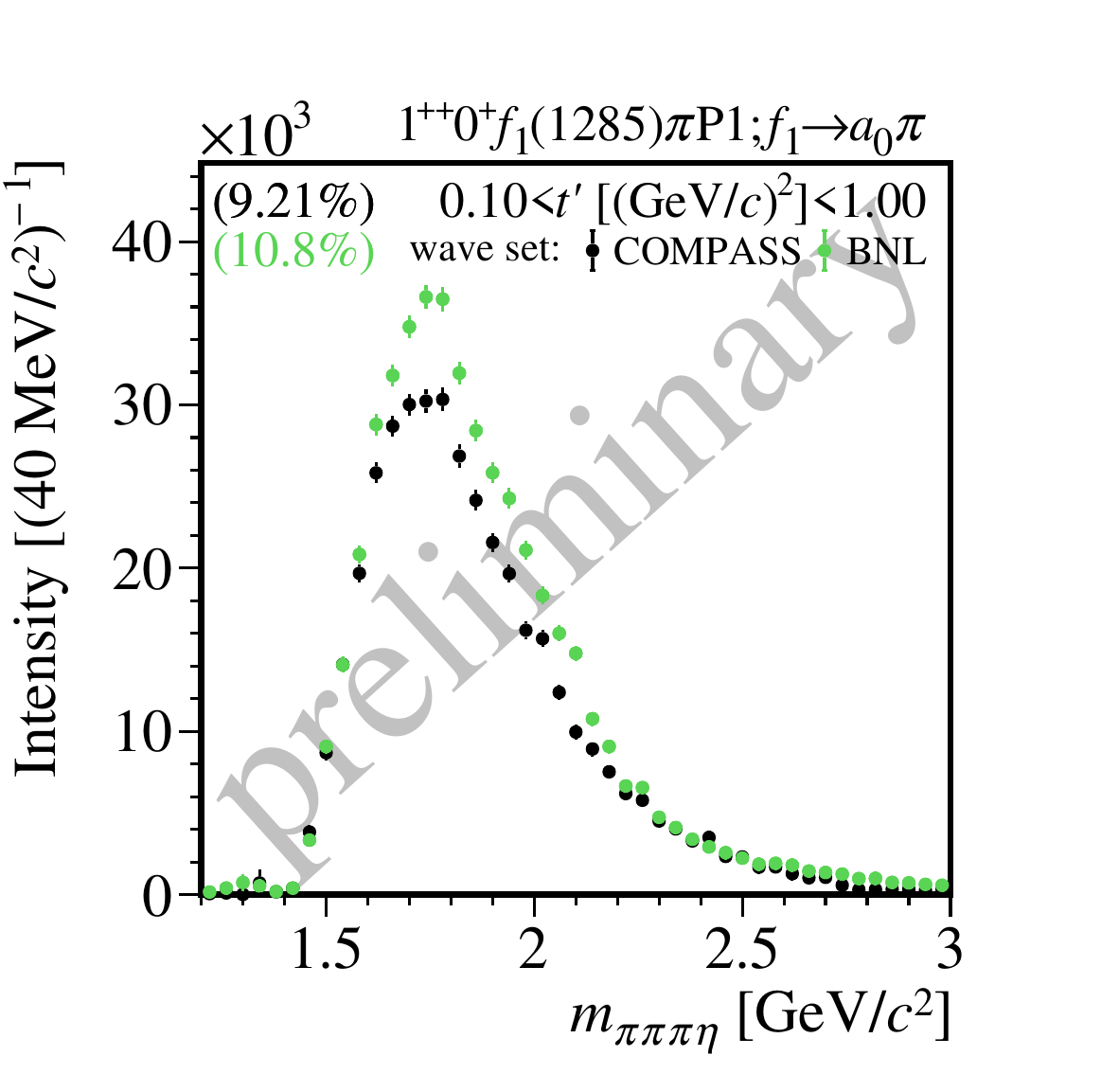}}}
        \qquad
        \caption{$t^\prime$-summed intensity distributions of the $1^{-+}1^+f_1(1285)\pi^-S1$ (a) and the $1^{++}0^+f_1(1285)\pi^-P1$ (c) waves. (b) The relative phase between these two waves. The black data points are our main results using our wave-set and the green data points is a fit to our data using the wave-set, which was used by the analysis in Ref.~\cite{BNL_F1}.}
        \label{fig:BNLcomparison}
    \end{figure}

For comparison, we examine the results of the BNL E852 collaboration, which studied the same system using a beam momentum of \SI{18}{GeV/c} \cite{BNL_F1}. Despite having roughly an order of magnitude fewer events and no $t^\prime$ binning, their analysis remains the most comprehensive study of this system to date. Their key finding was the presence of two resonances in the $1^{-+}1^+f_1(1285)\pi S1$-wave, identified as the $\pi_1(1600)$ and a possible excited $\pi_1$ state just above \SI{2}{GeV/c^2}. As a systematic cross-check, we apply the BNL model to our data with our $t^\prime$-binning and compare the distributions as shown in Fig.~\ref{fig:BNLcomparison}. %Fig.~\ref{fig:BNLcomparison} shows a comparison of the intensity distributions for the $1^{-+}1^+f_1(1285)\pi S1$-wave (a) and the $1^{++}0^+f_1(1285)\pi P1$-wave (c), as well as their relative phase difference (b). Our main results are shown as black points, while the outcome using the BNL model is shown in green. 

The approximately constant phase below \SI{2}{GeV/c^2} can be attributed to the presence of the $\pi_1(1600)$ and $a_1(1640)$ resonances with similar parameters. However, significant model dependence is observed in the intensity distributions: the exotic-wave intensity nearly vanishes in our model, and the pronounced negative phase motion above \SI{2}{GeV/c^2} strongly supports a resonance in the $1^{++}$-wave and not in the spin-exotic $1^{-+}$-wave. These observations contradict the claim of an excited $\pi_1$ state in \cite{BNL_F1}. Such model dependence has been observed previously in $3\pi$ analysis of COMPASS data~\cite{ModelDep}. It highlights the importance of systematic wave-selection studies. %Previous COMPASS analyses, such as that of the $(3\pi)^-$ system, have already highlighted and resolved artifacts stemming from such model dependencies \cite{ModelDep}. 

% Notes 
    % \begin{itemize}
    %     \item Introduce BNL analysis
    %     \item Lower beam energy, less events, no $t^\prime$-binning
    %     \item Own wave-set, which we use to fit our data
    %     \item Comment on excited $\pi_1 state$ 
    % \end{itemize}

% \newpage
% Improved version
\begin{wrapfigure}{r}{0.4\textwidth}
  % \vspace{-10pt} % Optional: pull the figure up slightly
  \centering
  \includegraphics[width=0.4\textwidth]{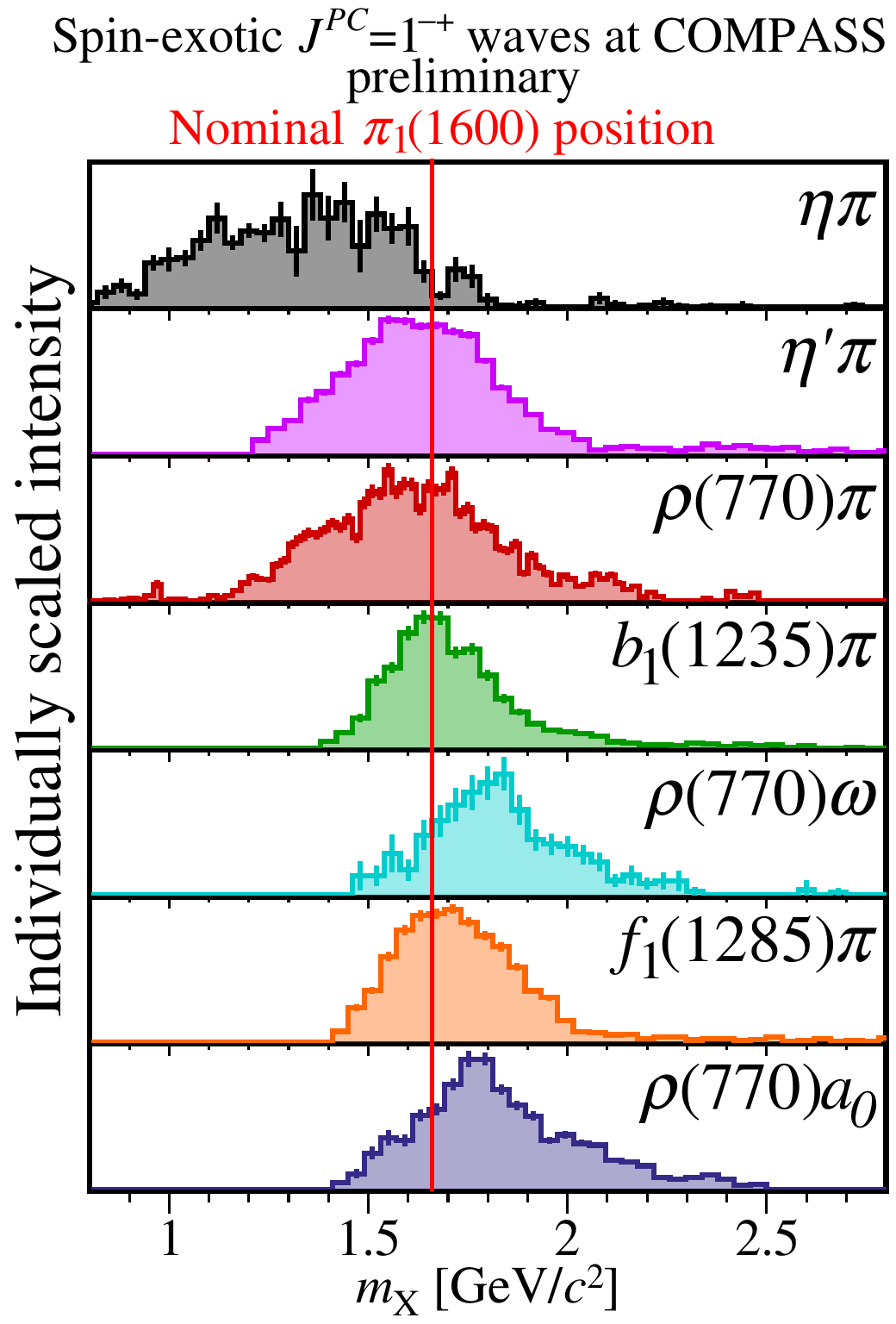} % Replace with your figure
  \caption{Intensities of the spin-exotic waves with (potential) $\pi_1(1600)$ signal in COMPASS data.}
  \label{fig:SpinExotics}
  % \vspace{-20pt} % Optional: reduce space after the figure
\end{wrapfigure}
To date, seven partial waves with spin-exotic quantum numbers and resonance-like signals from four different final states have been observed in COMPASS data. The signal in those waves, that have already been analyzed in the scope of a resonance-model fit, could be attributed to the $\pi_1(1600)$. The intensity distributions of all seven waves are shown in Fig.~\ref{fig:SpinExotics}. Three of these waves ($\eta^\prime\pi$, $f_1(1285)\pi$, $\rho(770)a_0$) are extracted from the analysis of the $\pi^-\pi^+\pi^-\eta$ system. It is important to note that these distributions may contain both resonant signals and coherent background contributions. The variation in peak positions and shapes arises from different signal strengths and interferences with the non-resonant background, underscoring the necessity of accurate mass-dependent modeling. To prove the hybrid meson interpretation of the $\pi_1(1600)$, determining relative branching ratios of its dominant decay channels is crucial. However, this task becomes experimentally challenging when the corresponding partial waves belong to different systems. In this context, the $\pi^-\pi^+\pi^-\eta$ system offers a unique opportunity to study several decay modes simultaneously, making it an ideal system to pin down the hybrid nature of the $\pi_1(1600)$. 

% Notes:
    % \begin{itemize}
    %     \item Discuss 7 spin-exotic waves at COMPASS with potential $\pi_1$-signal
    %     \item highlight 3 of them in 3pieta, two predicted to be significant, one not predicted
    %     \item Comment on importance of branching ration extraction to nail down hybrid hypothesis
    % \end{itemize}

% \nopagebreak
% \vspace{-2ex}
\section{Conclusions}
We have presented the first partial-wave decomposition of the diffractively produced $\pi\pi\pi\eta$ final state using the world’s largest data set collected by COMPASS. Resonance-like structures are observed in more than 15 partial waves, opening the door to a detailed study of the light isovector excitation spectrum, including previously unexplored decay channels. Among these is the spin-exotic $\pi_1(1600)$. We find no clear evidence for an excited $\pi_1$ state as previously reported in \cite{BNL_F1}, and we highlight the model dependence of such interpretations. The next step in the analysis will be the application of a resonance-model fit to extract the underlying resonance parameters and test the hybrid meson hypothesis of the $\pi_1(1600)$ more rigorously.

% Notes
% \begin{itemize}
%     \item First PWD of 3pieta with $t^\prime$-binning
%     \item Main sectors and channels
%     \item Many resoance-like signals
%     \item Three potential decay-channles of $\pi_1(1600)$
%     \item No signature of excited $\pi_1$. BNL distribution can be reproduced with other wave-set -> excited $\pi_1$ might be a model artifact
%     \item COMPASS measured so far 7 waves with spin-exotic quantum number 1-+. Three of then end up in 3pieta. Perfect playground to estimate branching ratios and nail down hybrid hypothesis. 
% \end{itemize}


\begin{thebibliography}{99}
\bibitem{COMPASS_hadron}
P. Abbon \textit{et al.} (COMPASS Collaboration), Nucl. Instrum. Meth. A \textbf{779}, 69 (2015).
\bibitem{LightIsoVector}
M. Aghasyan \textit{et al.} (COMPASS Collaboration), Phys. Rev. D \textbf{98}, 092003 (2018).
\bibitem{OddAndEven}
C. Adolph \textit{et al.} (COMPASS Collaboration), Phys. Lett. B \textbf{740}, 303 (2015).
\bibitem{OmegaPiPi}
P. Haas (for the COMPASS Collaboration), Nuovo Cim.C 47 (2024) 4, 150.
\bibitem{KsK}
F. M. Kaspar and J. Beckers (for the COMPASS Collaboration), EPJ Web \textbf{291}, 02014 (2024).
\bibitem{triangle}
G.D. Alexeev \textit{et al.} (COMPASS Collaboration), Phys. Rev. Lett. \textbf{127}, 082501 (2021).
\bibitem{poleofpione}
A. Rodas \textit{et al.} (JPAC Collaboration), Phys. Rev. Lett. \textbf{122}, 042002 (2019).
\bibitem{dudek}
A. J. Woss \textit{et al.} (Hadron Spectrum Collaboration), Phys. Rev. D \textbf{103}, 054502 (2021).
\bibitem{BNL_F1}
J. Kuhn \textit{et al.} (E852 Collaboration), Phys. Lett. B \textbf{595}, 109 (2004). %\url{https://doi.org/10.1016/j.physletb.2004.05.032}
\bibitem{Kubis}
T. Isken \textit{et al.}, Eur. Phys. J. C \textbf{77}, 489 (2017).
\bibitem{Ketzer}
B. Ketzer \textit{et al.}, Prog. Part. Nucl. Phys., Volume \textbf{113}, 103755 (2020).

\bibitem{ModelDep}
G. D. Alexeev \textit{et al.} (COMPASS Collaboration), Phys. Rev. D \textbf{105}, 012005 (2022).

\end{thebibliography}
\end{document}